\documentclass[aps,prb,superscriptaddress,footinbib,onecolumn,preprint]{revtex4} 
\usepackage{graphicx}
\usepackage{dsfont}
\usepackage{version} 
\usepackage{color}
\usepackage{amsmath}
\usepackage{mathtools}
\usepackage{braket}
\usepackage{relsize}
\usepackage{bm}
\usepackage{placeins}
\usepackage[separate-uncertainty=true]{siunitx}
\usepackage{upgreek}
\usepackage{colortbl}
\usepackage{gensymb}
\usepackage{graphicx}
\usepackage{dcolumn}
\usepackage{bm}
\usepackage{bigints}
\usepackage{bibunits}
\usepackage{xcolor}  
\usepackage{hyperref} 
\hypersetup{
    bookmarks=true,         
    unicode=false,          
    pdftoolbar=true,        
    pdfmenubar=true,        
    pdffitwindow=false,     
    pdfstartview={FitH},    
    pdftitle={Dynamical Exploration of Amplitude Bistability in Engineered Quantum Systems},    
    pdfauthor={Andreas Angerer},     
    pdfkeywords={keyword1} {key2} {key3}, 
    pdfnewwindow=true,      
    colorlinks=true,       
    linkcolor=red,          
    citecolor=blue,        
    filecolor=magenta,      
    urlcolor=cyan,           
}

\begin{document}

\title{Dynamical Exploration of Amplitude Bistability in Engineered Quantum Systems} 

\author{Andreas Angerer$^\S$}
\email{andreas.angerer@tuwien.ac.at}
\affiliation{Vienna Center for Quantum Science and Technology, Atominstitut, TU Wien, Stadionallee 2, 1020 Vienna, Austria}
\affiliation{Zentrum f\"ur Mikro- und Nanostrukturen, TU Wien, Floragasse 7, 1040 Vienna, Austria}

\author{Stefan Putz$^\S$}
\affiliation{Vienna Center for Quantum Science and Technology, Atominstitut, TU Wien, Stadionallee 2, 1020 Vienna, Austria}
\affiliation{Zentrum f\"ur Mikro- und Nanostrukturen, TU Wien, Floragasse 7, 1040 Vienna, Austria}
\affiliation{Department of Physics, Princeton University, Princeton, New Jersey 08544, USA}

\author{Dmitry O. Krimer}
\affiliation{Institute for Theoretical Physics, TU Wien, Wiedner Hauptstra{\ss}e 8-10/136, 1040 Vienna, Austria}

\author{Thomas Astner}
\affiliation{Vienna Center for Quantum Science and Technology, Atominstitut, TU Wien, Stadionallee 2, 1020 Vienna, Austria}
\affiliation{Zentrum f\"ur Mikro- und Nanostrukturen, TU Wien, Floragasse 7, 1040 Vienna, Austria}

\author{Matthias Zens}
\affiliation{Institute for Theoretical Physics, TU Wien, Wiedner Hauptstra{\ss}e 8-10/136, 1040 Vienna, Austria}

\author{Ralph Glattauer}   
\affiliation{Vienna Center for Quantum Science and Technology, Atominstitut, TU Wien, Stadionallee 2, 1020 Vienna, Austria}

\author{Kirill Streltsov}   
\affiliation{Vienna Center for Quantum Science and Technology, Atominstitut, TU Wien, Stadionallee 2, 1020 Vienna, Austria}

\author{William J. Munro}   
\affiliation{NTT Basic Research Laboratories, 3-1 Morinosato-Wakamiya, Atsugi, Kanagawa 243-0198, Japan}

\author{Kae Nemoto}   
\affiliation{National Institute of Informatics, 2-1-2 Hitotsubashi, Chiyoda-ku, Tokyo 101-8430, Japan}

\author{Stefan Rotter}   
\affiliation{Institute for Theoretical Physics, TU Wien, Wiedner Hauptstra{\ss}e 8-10/136, 1040 Vienna, Austria}

\author{J\"org Schmiedmayer}
\affiliation{Vienna Center for Quantum Science and Technology, Atominstitut, TU Wien, Stadionallee 2, 1020 Vienna, Austria}

\author{Johannes Majer}
\affiliation{Vienna Center for Quantum Science and Technology, Atominstitut, TU Wien, Stadionallee 2, 1020 Vienna, Austria}
\affiliation{Zentrum f\"ur Mikro- und Nanostrukturen, TU Wien, Floragasse 7, 1040 Vienna, Austria}

\let\thefootnote\relax\footnotetext{{$^\S$ These authors contributed equally to this work}}
\date{\today}

\begin{abstract}
\bf{
Nonlinear systems, whose outputs are not directly proportional to their inputs, are well known to exhibit many interesting and important phenomena which have  profoundly changed our  technological landscape over the last 50 years. Recently the ability to engineer quantum metamaterials through hybridisation \cite{imamoglu_cavity_2009,kubo_hybrid_2011,amsuss_cavity_2011,schuster_high-cooperativity_2010,xiang_hybrid_2013,kurizki_quantum_2015} has allowed to explore these nonlinear effects in systems with no natural analogue. Here we investigate amplitude bistability \cite{rempe_optical_1991}, which is one of the most fundamental nonlinear phenomena, in a hybrid system composed of a superconducting resonator inductively coupled to an ensemble of nitrogen-vacancy centres \cite{jelezko_observation_2004,doherty_nitrogen-vacancy_2013}. One of the exciting properties of this spin system is its extremely long spin life-time, more than ten orders of magnitude longer than other relevant timescales of the hybrid system \cite{jarmola_temperature-_2012}. This allows us to dynamically explore this nonlinear regime of cavity quantum electrodynamics (cQED) and demonstrate a critical slowing down of the cavity population on the order of several tens of thousands of seconds - a timescale much longer than observed so far for this effect \cite{mitschke_transients_1983,garmire_transient_1979}. Our results provide the foundation for future quantum technologies based on nonlinear phenomena.}
\end{abstract}
\maketitle

In nature, most physical systems are inherently nonlinear, giving rise to effects such as bistability \cite{gibbs_optical_1979}, chaos \cite{benza_symmetry_1987}, solitons \cite{mollenauer_experimental_1980}, superradiance \cite{dicke_coherence_1954} and often appear counter-intuitive when contrasted with much simpler linear systems.   Amplitude bistability,  one of the basic nonlinear phenomena (nowadays commonly used in optical switches \cite{szoke_bistable_1969}) has been extensively investigated both theoretically \cite{casteels_power_2016,bonifacio_transient_1978,sawant_optical-bistability-enabled_2016,dombi_optical_2013,martin_extreme_2011} and experimentally \cite{rempe_optical_1991,kimble_strong_1998}. It occurs in any medium where strong nonlinearities  in the interaction between a radiation field and a polarisable medium such as spins exist. The nonlinearity in such systems arises  from the two-level nature of the atoms coupled to the cavity mode but only shows up when driven beyond the single excitation regime. For a strong coupling between the spin system and the cavity mode a first order phase transition between a saturated, disordered and a de-excited, ordered ground state occurs \cite{bonifacio_cooperative_1976}. The coupled system switches between these two branches and shows a hysteresis depending on the history of the system.

The usual cQED demonstrations use atoms or trapped ions coupled to optical light fields to investigate these nonlinear effects but the short atomic life-times have made it difficult  to truly observe the temporal dynamics of amplitude bistability \cite{mitschke_transients_1983,garmire_transient_1979}, and restricted previous studies to the steady state behaviour.  In contrast emerging quantum engineering through hybridisation allows us to create systems with extremely long-lived emitters, making it possible to observe the bistable system during evolution. In this letter, we report on the observation of amplitude bistability in  a cQED system composed of a superconducting resonator coupled to a long-lived electron spin ensemble formed from artificial atoms (NV$^-$ centres in diamond \cite{jelezko_observation_2004,doherty_nitrogen-vacancy_2013}). Their extremely long life-times allow us to study the temporal behaviour of the presented effect, a regime experimentally not accessible so far.


An ensemble of spins in a cavity is characterised by the three quantities polarisation, inversion and the cavity amplitude. They can be derived from the driven Tavis-Cummings Hamiltonian \cite{tavis_exact_1968} for $N$ spins under a rotating wave approximation,
\begin{eqnarray}
\mathcal{H}=\hbar\omega_c\boldsymbol a^{\dagger}\boldsymbol a+\frac{\hbar}{2}\sum_{j=1}^N\omega_j \boldsymbol \sigma_j^z+i\hbar \sum_{j=1}^Ng_j \left(\boldsymbol \sigma^-_{j}\boldsymbol a^{\dagger}-\boldsymbol \sigma^+_j \boldsymbol a\right)\nonumber\\
+i\hbar\left(\eta \boldsymbol a^{\dagger}e^{-i\omega_p t}-h.c.\right),\,\,\,\,\,\,\,\,
\label{eq:Hamilt_fun}
\end{eqnarray}
with $\boldsymbol a^{\dagger}$, $\boldsymbol a$ as the creation and annihilation operators for the cavity mode of frequency $\omega_c$ and  $\boldsymbol \sigma_j^z$, $\boldsymbol\sigma_{j}^+$, $\boldsymbol\sigma_{j}^-$ as the spin inversion, raising and lowering operators for the $j$-th spin of frequency $\omega_j$ coupled to the cavity with a single spin coupling strength $g_j$. The last term accounts for an external cavity drive with field amplitude $\eta$ and frequency $\omega_p$.

Using a mean-field approximation, valid in the limit of very large spin ensembles, $\braket{\boldsymbol a^\dagger \boldsymbol \sigma_-}\approx a^{\dagger}\sigma_-$ (from now on unbolded symbols will be used for the expectation values),  we derive a set of first order differential equations, formally equivalent to the well known Maxwell-Bloch equations \cite{bonifacio_theory_1982} as
\begin{eqnarray}
&\dot{a} =  -\kappa a + \sum_j\nolimits g_j  \sigma^-_j+\eta \nonumber\\
&\dot{\sigma}^-_j=  -\left(\gamma_\perp+i\Theta_j\right)  \sigma^-_j +g_j\sigma^z_ja\\
&\dot{\sigma}^z_j=  -\gamma_\parallel\left(1+\sigma^z_j\right)-2 g_j\left(  \sigma^-_j a^\dag+\sigma^+_j a\right)\nonumber,
\label{eq:maxwellbloch}
\end{eqnarray}
with cavity dissipation rate $\kappa$, transversal spin relaxation rate $\gamma_{\perp}=1/{T_2}$ and longitudinal spin relaxation rate $\gamma_{\parallel}=1/{T_1}$. The relaxation rates are ordered as $\kappa >  \gamma_{\perp} \gg \gamma_{\parallel}$, such that the spin inversion is by far the slowest process. $\Theta_j$ are frequency detunings with respect to the ensemble central frequency to account for inhomogeneous broadening. Setting the time derivatives to zero we obtain the steady state of this system as
\begin{eqnarray} \label{eq:a2}
\left|a\right|^2\!=\!\frac{\eta^2}{\kappa^2}\left(1-\sum_j\nolimits C_j\sigma_j^z\right)^{-2}\!\!\!,\,\,\,\,
\sigma_j^z\!=\!-\left(1+\frac{4  g_{j}^2 |a|^2\gamma_\perp}{\gamma_\parallel(\gamma_\perp^2+\Theta_j^2 )}\right)^{-1},
\end{eqnarray}%
\normalsize where the dimensionless parameter ${C_j= g_j^2/\left[\kappa\gamma_\perp\left(1+\Theta_j^2/\gamma_\perp^2\right)\right]}$ is the single spin cooperativity. The collective system cooperativity is given accordingly by $C_\mathrm{coll}=\sum_j C_j$.

We can classify the expected system phase transition by deriving a solution for the time dependent cavity amplitude $|a(t)|^2$. The difference in dissipation rates allows us to adiabatically eliminate the $a$ and ${\sigma}^-_j$ variables \cite{lugiato_nonlinear_2015}, which results in a first order  differential equation for the intra-cavity intensity. For the giant $S_z=\sum _j\sigma^z_j$ spin in resonance ($\Theta_j=0$) with the cavity mode it can be written as
\begin{eqnarray}
\label{eq:adiabatic} \frac{ \mathrm d |a|^2}{\mathrm d t}\!&=&\! -\frac{8 C_\mathrm{{coll}} \kappa^2|a|^5}{\eta}\!+8 C_\mathrm{{coll}} \kappa |a|^4\!-\!\frac{2 \kappa\gamma_\parallel}{\eta}\left(1\!+\!C_\mathrm{{coll}}\right)|a|^3 \nonumber \\
&& \;\;\;\;\;\;\;\;\;\;\;\;\;\;\;+2\gamma_\parallel|a|^2.\!\!\!\!\!
\end{eqnarray}
For our typical system parameters this equation predicts a first order phase transition that connects a strongly driven branch and a weakly driven branch, with hysteresis and two saddle-node bifurications  \cite{strogatz_nonlinear_2001} at which a jump occurs between both branches. 

\begin{figure}[ht]
\includegraphics[width=0.5\linewidth]{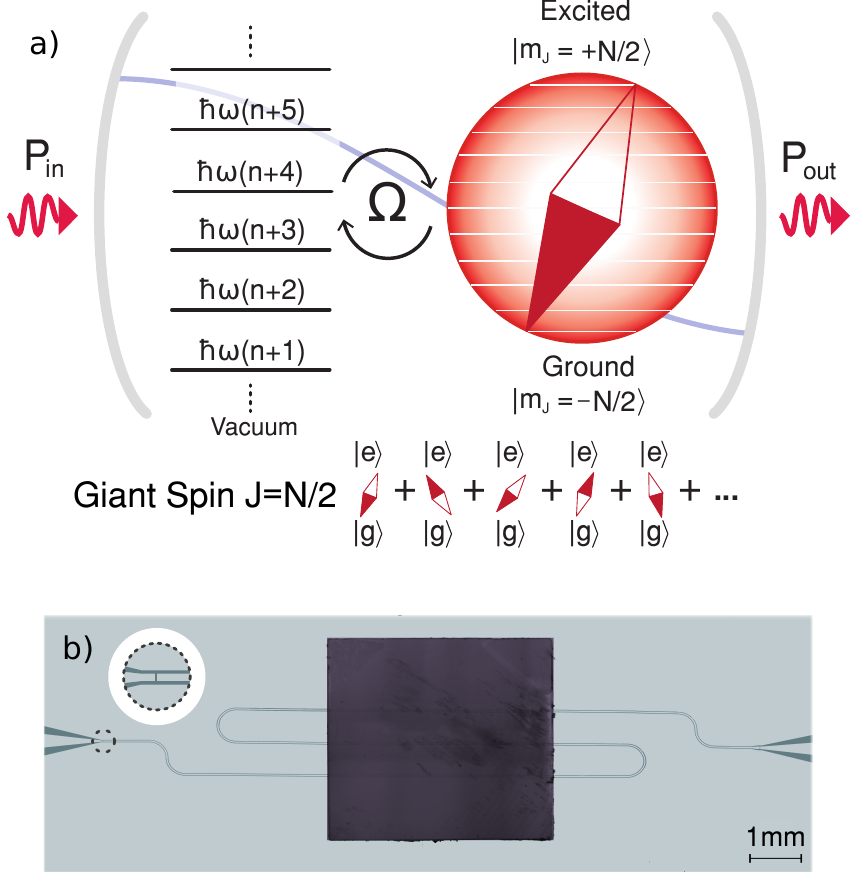}
\caption{\textbf{a, }Schematic illustration of our experimental setup in which an ensemble of spins (described as an effective giant spin) is inductively coupled (with a coupling rate $\Omega$) to the cavity mode. The nonlinearity stems from the anharmonicity of this coupled spin when driven beyond its linear regime, which we probe through the transmission $|\mathrm T|^2=\mathrm{P}_\mathrm{out}/\mathrm{P}_\mathrm{in}$ of the hybrid system. \textbf{b,} Photograph of the system consisting of a superconducting transmission line cavity with an enhanced neutron-irradiated diamond on top of it, containing a large ensemble of NV-spins (black).   Two coupling capacitors provide the necessary boundary conditions for the microwave radiation. }
 \label{fig:setup}
\end{figure}

Our hybrid system is illustrated in Fig. \ref{fig:setup} and is composed of an electron spin ensemble formed by NV centres in diamond, loaded onto a superconducting $\lambda/2$-resonator. To thermally polarise the $N\approx10^{12}$ electron spins to their ground state ($\geq$99 \%) we put the system in a dilution refrigerator at \SI{25}{mK}. Each electron spin has a zero field splitting of $D/2\pi \sim$ \SI{2.878}{GHz} and average coupling rate of $g_0/2\pi\approx \SI{12}{Hz}$ to the cavity mode.  
The  superconducting resonator has a cavity linewidth of $\kappa/2\pi$=\SI{440\pm10}{kHz} (HWHM) with a fundamental resonance frequency at $\omega_c/2\pi$=\SI{2.691}{GHz} and a loaded quality factor of $Q=3300$. An external microwave field with frequency $\omega_p$ is used to probe our hybrid system. 

\begin{figure}[ht!]
\includegraphics[width=0.45\linewidth]{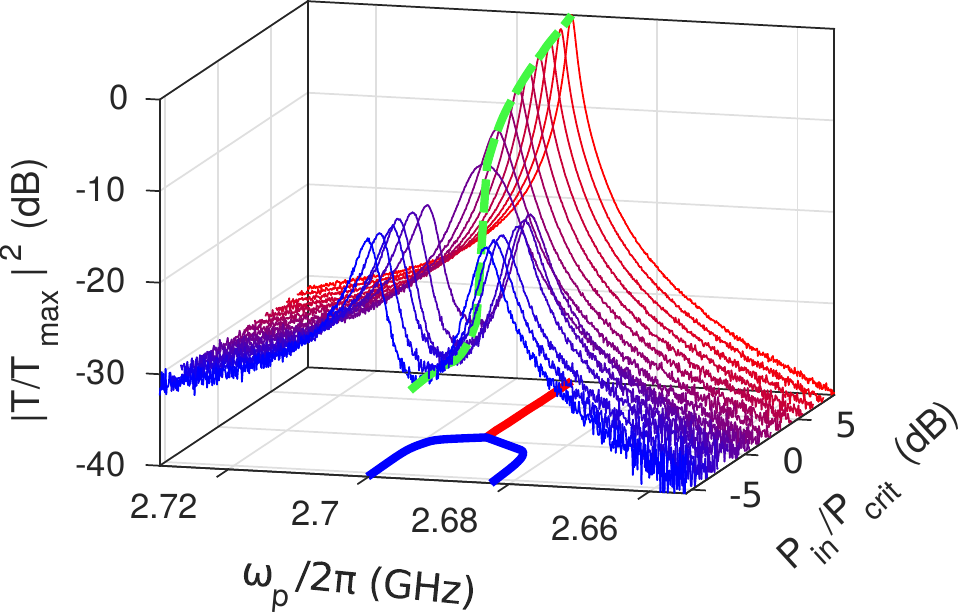}
\caption[]{Evolution of the transmission spectrum for different input drive powers $\mathrm{P}_\mathrm{in}$. In the linear regime the Rabi-splitting is clearly observable. For a drive power $\mathrm{P}_\mathrm{in}\approx\mathrm{P}_\mathrm{crit}$, the spin system starts to bleach and decouples from the cavity. For input drives   $\mathrm{P}_\mathrm{in}\gg\mathrm{P}_\mathrm{crit}$ we observe the bare cavity transmission function. This behaviour can be seen from the projection of the observed maximum transmission peaks on the $xy$-plane. When driving the system resonantly (see green dashed line) and with a cooperativity large enough, operating between these two regimes exhibits amplitude bistability.} \label{fig:rabi}
\end{figure}

First we search for the  steady state bistable behaviour by measuring the transmitted intensities  through the cavity defined by $|\mathrm T|^2=\mathrm{P}_\mathrm{out}/\mathrm{P}_\mathrm{in}$ as a function of the input drive intensity $\mathrm{P}_\mathrm{in} \approx \eta^2/\kappa$ and outgoing intensity $\mathrm{P}_\mathrm{out}\approx\left|a\right|^2\kappa$. The drive power is raised in a stepwise manner, slow enough to allow the system to reach a steady state for each stimulus $\mathrm{P}_\mathrm{in}$. For small excitations the intra-cavity intensity is not sufficient to saturate the spin ensemble ($\sigma_j^z\sim-1$) and is thus given by $\left|a\right|^2\approx \frac{\eta^2}{\kappa^2}\frac{1}{(1+ C_\mathrm{coll})^2}$. As the power level increases, the cavity field bleaches the spins ($\sigma_j^z\approx \sigma_j^-\sim0$) such that the Rabi-splitting vanishes and the spin system decouples from the cavity (Fig.~\ref{fig:rabi}). The intra-cavity intensity $\left|a\right|^2\approx \frac{\eta^2}{\kappa^2}$ is that of an empty cavity from which spins are completely decoupled. We identify the power level for which the transition between both cases occur as the critical drive $\mathrm{P}_\mathrm{crit}$.

\begin{figure*}[bhtb]
\includegraphics[width=1\linewidth]{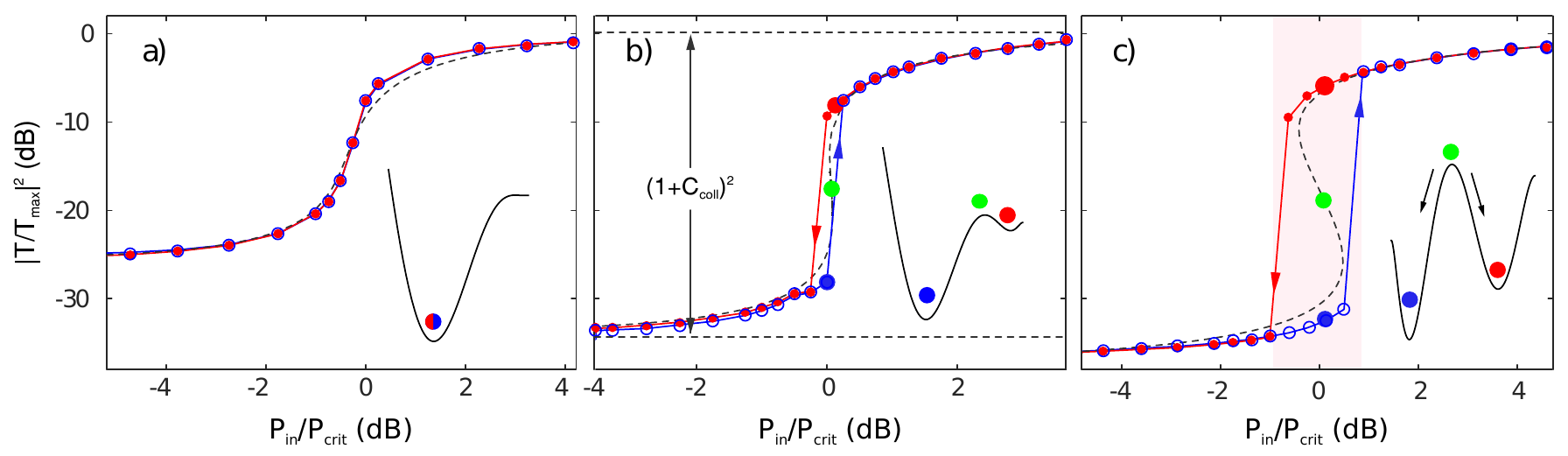}
\caption{Steady state bistability transmission measurements through the cavity for our hybrid system as a function of increasing (\textit{blue}) and decreasing (\textit{red}) input power $\mathrm{P}_\mathrm{in}$. In \textbf{a,} we plot the transmission measurement with a cooperativity $C_\mathrm{coll}\approx18$ and cavity linewidth of $\kappa/2\pi=\SI{1.2}{MHz}$ using two subensembles in resonance with the cavity.  \textbf{b,} The same transmission measurement with $C_\mathrm{coll}\approx49$ and $\kappa/2\pi=\SI{0.44}{MHz}$. A small bistable area is visible where no steady states exist and the system jumps from one steady state to the next. \textbf{c,} The same measurement as in \textbf{a} with an increased cooperativity of $C_\mathrm{coll}\approx78$ (by using all four NV subensembles in resonance with the cavity), again with $\kappa/2\pi=\SI{0.44}{MHz}$. A large $\approx\SI{2}{dB}$ clear bistable regime is seen. The dashed lines in all subfigures correspond to the predictions from our model of a saddle-node bifurication with two fixed points at the turning points, the lower one stable (blue solid dot) and the upper one unstable (red solid dot). For all three measurements also a sketch of the corresponding potential, derived from solving Eq.~(\ref{eq:adiabatic}), is drawn that shows the occurence of either one or two stable solutions.}
\label{fig:bistab}
\end{figure*}

This nonlinear saturation behaviour is a necessary precursor to the observation of amplitude bistability. However, whether or not this is observable in the experiment is determined by the system's collective cooperativity. This is apparent from Eq.~(\ref{eq:a2}), where larger cooperativity values result in stronger nonlinearity and thus a larger phase seperation. In Fig.~\ref{fig:bistab} we show steady state bistability measurements for three cooperativity values $C_\mathrm{coll}=18,49,78$. The lowest value $C_\mathrm{coll}=18$ does not show bistability (Fig.~\ref{fig:bistab}a), but increasing the cooperativity to  $C_\mathrm{coll}=49$ (see Methods) allows us to observe the  first signs of bistable behaviour (see Fig.~\ref{fig:bistab}b). Increasing the cooperativity further to $C_\mathrm{coll}\approx78$ shows clear amplitude bistability (Fig.~\ref{fig:bistab}c) within a 2dB range. This steady state bistability behaviour is well reproduced by a full numerical simulation taking inhomogeneouse broadening into account (dashed lines in Fig.~\ref{fig:bistab}a-c). 

\begin{figure*}[bht]
\includegraphics[width=1\linewidth]{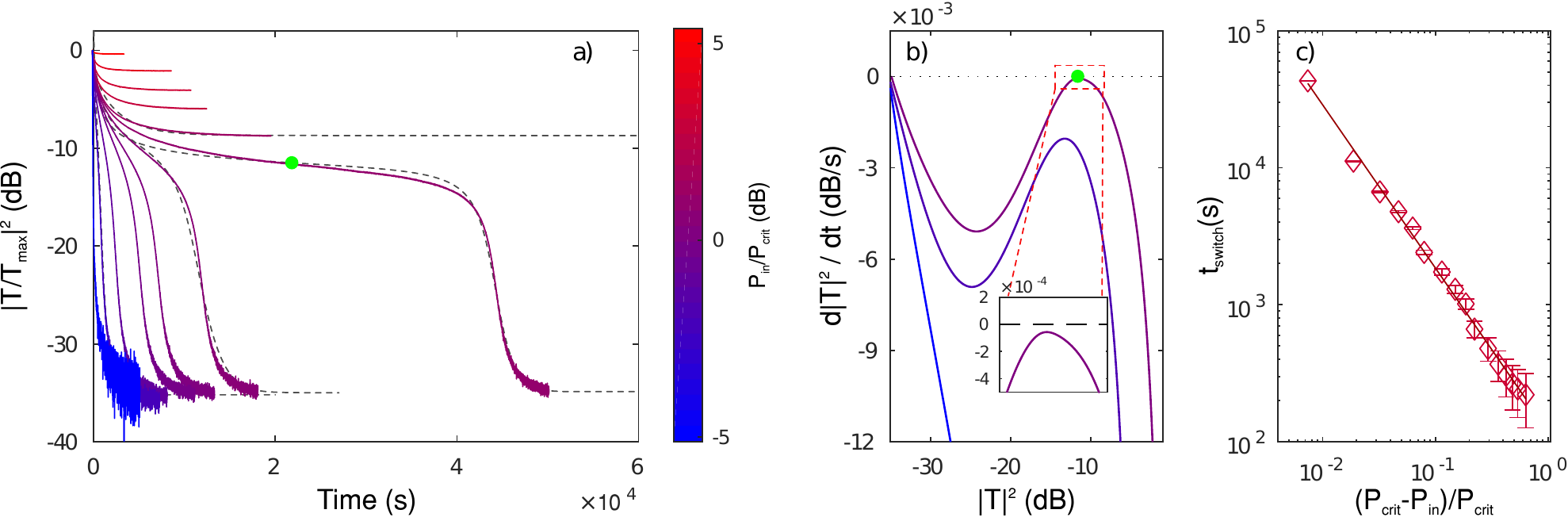}
\caption{ Quench dynamics of high cooperativity $C_\mathrm{coll}\approx78$ configuration. In \textbf {a,} the intra-cavity intensity $|\mathrm{T}|^2/|\mathrm{T}_{\rm max}|^2$ is plotted over time for different drive intensities where we observe that the time to reach a steady state depends on the input drive intensity.  For drive intensities larger than a critical drive value $\mathrm{P}_\mathrm{crit}$ (defined as the power where the system undergoes the phase transition from the lower to the upper branch, see Fig.~\ref{fig:bistab})  the spin system remains saturated, whereas for drive intensities much smaller than $\mathrm{P}_\mathrm{crit}$ the spin system starts to collectively decay into its ground state. Close to the critical drive $\mathrm{P}_\mathrm{crit}$ this timescale is extremely prolonged and approaches \SI{4e4}{s}. The dashed lines correspond to predictions from our model. In  \textbf {b,} we show the phase diagram as $\mathrm{d}|\mathrm{T}|^2/\mathrm{d}t$ over $|\mathrm{T}|^2$, for the decay towards a steady state (black dotted line) for different input drives $\mathrm{P}_\mathrm{in}$. For drive powers close to the critical drive the derivative approaches zero and the decay becomes much smaller compared to drive powers larger and smaller than the critical drive. The switching time  between the upper and lower branch for different input drives are shown in \textbf {c,}. We define the switching time as the inverse of the smallest decay rate for a given curve (green dot in (a-b)). Close to the critical drive the switching time diverges, and the time to reach a steady state becomes arbitrarily long. The solid red line is a fitting function of the form $t_\mathrm{switch}\sim|\mathrm{P}_\mathrm{in}-\mathrm{P}_\mathrm{crit}|^{-\alpha}$ (with $\alpha=1.2$).}
\label{fig:quench}
\end{figure*}

Given such clear evidence of amplitude bistability, we focus next on the temporal behaviour of the hybrid system using quench dynamic measurements.  Here the system is prepared in an initial steady state in which the spin ensemble is completely saturated and constantly driven with $\mathrm{P}_\mathrm{in}\gg\mathrm{P}_\mathrm{crit}$. The drive power is non-adiabatically switched to a lower drive level and the system transmission is monitored. We repeat this measurement several times, always preparing the system in the same initial state, but switching to different lower drive powers. When the system is driven close to the bifurication point ($\mathrm{P}_\mathrm{in} \approx\mathrm{P}_\mathrm{crit}$) the timescales needed to settle in a stationary state become as long as \SI{40e3}{\second}, as depicted in Fig.~\ref{fig:quench}. This is known as critical slowing down \cite{grynberg_critical_1983,garmire_transient_1979}. 

The behaviour can be linked to our model given in Eq.~(\ref{eq:adiabatic}) which predicts that our system features two fixed points at which a saddle-node bifurication occurs. A branch with no stable solutions connects two stable branches, one weakly driven, with  de-excited and ordered spins, and the other one strongly driven, with unordered and saturated spins (see Fig.~\ref{fig:bistab}c).  Starting in the strongly driven upper branch with a large intra-cavity intensity, the saturated spin system ($\sigma_j^z\approx \sigma_j^-\sim0$) is entirely decoupled from the cavity. This prohibits collective decay into the cavity leaving the extremely small longitudinal relaxation rate $\gamma_\parallel$ as the dominant decay channel. In the opposite limit of drive intensities much smaller than the critical value, the spin system starts to collectively decay through the cavity early on with a rate much larger than $\gamma_\parallel$ and no critical slowing down is observed. 

At the critical drive intensity, the saturation due to the drive and decay of the spin system are equal and opposite in effect allowing an everlasting decay to occur without the system reaching its fixed point.  If however, the drive intensity is slightly smaller than the critical drive, we observe a critical slowing down of the decay, but eventually the decay of the collectively decaying spin system becomes dominant which leads to a buildup of correlations in the system and  a faster decay of the spins and thus the cavity population.  This behaviour is shown in Fig.~\ref{fig:quench}a-c where close to a critical drive the system evolves towards the upper unstable fixed point, with a time derivative that can approach zero arbitrarily closely (inset in Fig.~\ref{fig:quench}b).  Small deviations from the critical drive lead to a speed up in decay until the system relaxes to a real steady state in the end. The time it takes to go from the upper to the lower branch diverges close to the critical drive according to $t_\mathrm{switch}\sim|\mathrm{P}_\mathrm{in}-\mathrm{P}_\mathrm{crit}|^{-\alpha}$, as shown in Fig.~\ref{fig:quench}c with $\alpha\sim 1.2$. This algebraic divergence is characteristic for nonlinear systems showing saddle-node bifurications  \cite{kuehn_scaling_2009}.  Comparing these experimental results with the full numerical solutions of Eqs.~(\ref{eq:maxwellbloch}) including inhomogeneous broadening, we observe excellent agreement (see Fig.~\ref{fig:quench}a).

To summarise, we have shown how a hybrid system composed of a superconducting resonator coupled to an electron spin ensemble in diamond can be used to explore amplitude bistability in new regimes of cQED, with unusual decay rates where the spin life-time is much smaller than other decay constants in the system.  We have demonstrated a critical slowing down of the cavity population on the order of eleven hours, a timescale several orders of magnitude longer than observed so far for this effect and many order of magnitude longer than other time scales associated with the system. Our experiment provides a foundation for the exploration of new nonlinear phenomena in  quantum metamaterials and future quantum technologies that may arise from it.

\textbf{~\\Acknowledgements}\\
We would like to thank Helmut Ritsch, Michael Trupke and Alberto Amo for helpful discussions. The experimental effort lead by J.M. has been supported by the Top-/Anschubfinanzierung grant of the TU Wien. S.P., A.A. and T.A. acknowledge support by the Austrian Science Fund (FWF) in the framework of the Doctoral School “Building Solids for Function” Project W1243. D.O.K. and S.R. acknowledge funding by the Austrian Science Fund (FWF) through the Spezialforschungsbereich (SFB) NextLite Project No. F49-P10. K.N. acknowledges support from the MEXT KAKENHI Grant-in-Aid for Scientific Research on Innovative Areas “Science of hybrid quantum systems” Grant No. 15H05870.

\textbf{~\\Author contributions}\\
S.P. and D.O.K. conceived the idea and S.P., J.S, and J.M. designed and set up the experiment. A.A., S.P., R.G., T.A., and K.S. carried out the measurements under the supervision of J.M.. D.O.K, M.Z., and S.R. devised the theoretical framework and provided the theoretical support for modelling the experiment. K.N. and W.J.M provided support for implementation of the theoretical methods. A.A. wrote the manuscript to which all authors suggested improvements.\\
\newpage

\newpage

\textbf{~\\Methods}\\
\small{
\\\textbf{Sample}\\
The spin system is realised by enhancing a type Ib high-pressure high-temperature (HPHT) diamond crystal containing an initial concentration of \SI{200}{ppm} nitrogen with a natural abundance of $\prescript{13}{}{\mathrm{C}}$ nuclear isotopes. We achieve a total density of $\approx \SI{6}{ppm}$ negatively charged NV centres by \SI{50}{h} of neutron irradiation with a fluence of \SI{5e17}{cm^{-2}} and annealing the crystal for \SI{3}{h} at \SI{900}{\celsius}. Excess nitrogen P1 centres ($S = 1/2$), uncharged $\mathrm{NV}^0$ centres, plus additional
lattice stress serves as the main source of inhomogeneous broadening, which exceeds decoherence due to the naturally abundant 1.1\% $\prescript{13}{}{\mathrm{C}}$ spin bath. The characteristics of the diamond crystal and NV ensemble were initially determined at room temperature using an optical confocal microscope. In the present experiment, the broadened spin ensemble was characterised by a q-Gaussian spectral line shape $\rho(\omega)$ with a linewidth of $\gamma_\mathrm{inh} = \SI{4.55}{MHz}$.\\
\\\textbf{Spin system}\\
The negatively charged $\mathrm{NV}^-$ centre is a paramagnetic point-defect centre in diamond with an electron spin $S=1$, consisting of a nitrogen atom replacing a carbon atom in the diamond lattice and an adjacent vacancy. The ground spin triplet can be described by a simplified Hamiltonian $H/h=D S_z^2+\mu_BS_z$, with $\mu=\SI{28}{MHz/mT}$ and a large zero-field splitting of $D/2\pi=\SI{2.87}{GHz}$. The splitting corresponds to a temperature of $D/hk_b=\SI{138}{mK}$, which allows to thermally polarise the spins in the ground state at the fridge base temperature of $\SI{25}{mK}$ with up to 99\% fidelity. Due to its crystallographic diamond structure four different NV subensembles, with equal abundance, pointing in the $[1,1,1]$ direction, exist. By applying $B\approx\SI{30}{mT}$ either with 0\degree or 45\degree relative to the $[1,0,0]$ direction in the NV-resonator plane, we can Zeeman tune four or two NV subensembles into resonance with the cavity mode. 
\\\\\textbf{Superconducting resonator}\\
The microwave cavity is loaded by placing the diamond sample on top of a $\lambda/2$-transmission line resonator. The superconducting microwave cavity is fabricated by optical lithography and reactive-ion etching of a \SI{200}{nm}-thick niobium film
sputtered on a \SI{330}{μm}-thick sapphire substrate. The loaded chip is hosted and bonded to a printed circuit board enclosed in a copper sarcophagus and connected to microwave transmission lines. The cavity exhibits a linewidth of $\kappa/2\pi=\SI{440}{kHz}$ which we can increase to $\kappa/2\pi=\SI{1.2}{MHz}$ by applying weak magnetic fields perpendicular to the resonator plane that partially quenches the superconducting material. The cavity is coupled to the environment such that the internal losses of the cavity are much smaller than the coupling losses ($\kappa_\mathrm{int} \ll \kappa_\mathrm{ext}$). This allows to approximate the total losses as $\kappa \approx \kappa_\mathrm{ext}$.
\\\\\textbf{Transmission measurements}\\
Transmission measurements are performed by recording the forward scattering parameter $|S_{21}|^2$ of the hybrid system using a standard vector network analyzer (Agilent E5071C). To perform steady state bistability measurements we probe the system first with increasing power levels. For each power level we monitor the transition until a steady state is reached. We increase the drive power until the steady state lies far in the high driving branch, after which the power is lowered again in a stepwise manner. We identify bistability if the system shows different steady states when driven with increasing and decreasing power. 
From Eq.~(\ref{eq:a2}), we immediately see that the asymptotic behaviour of the transmission intensity $|\mathrm{T}|^2=|a|^2\kappa^2/\eta^2$ is described by $|\mathrm{T_\mathrm{low}}|^2=\frac{1}{(1+ C_\mathrm{coll})^2}$ in the single excitation regime and by $|\mathrm{T_\mathrm{high}}|^2=1$ in the large excitation regime. The stroke between these two regimes is therefore only determined by the collective cooperativity  $C_\mathrm{coll}$ as defined in the main text. 
\\\\\textbf{Quench dynamic measurements}\\
Quench dynamics measurements are used to measure the temporal behaviour of the observed effect. For this we initialise the system in an initial state far in the large driving regime with a strong drive for several minutes. After this state preparation where the spin system is completely decoupled from the cavity, the drive power is non adiabatically switched to a lower drive, with transmission monitored for different target drive levels. We monitor the transmission until the time derivative of the transmission amplitude becomes smaller than an arbitrarily chosen threshold, which we then identify as our steady state.
\\\\\textbf{Transversal decay rate}\\
We employ Car-Purcell-Meiboom-Gill (CPMG)-like sequences to get an estimate for the spin-spin relaxation time ($T_2=1/\gamma_\perp$). The best achievable echo times in our experiment are $T_2 =\SI{4.8\pm1.6}{\micro s}$, which we identify as a lower bound for our relaxation times. The real spin-spin relaxation times are potentially longer, but misalignment of the external d.c. magnetic field with respect to the NV axis and a bath of excess electron and nuclear spins in the host material limits the echo time to times shorter than the real relaxation times.
\\\\\textbf{Longitudinal decay rate}\\
To get a value for $\gamma_\parallel$ we perform time resolved quench dynamic measurement in a regime with no bistability visible. For the limit $t\gg1/\gamma_\perp,1/\Theta_j,1/\kappa$  the decay of the cavity intensity follows a single exponential decay $\mathrm{e}^{-\alpha\gamma_\parallel t}$ with different values $\alpha$ for different drive intensities $\mathrm{P}_\mathrm{in}$.  By extracting these asymptotical decay constants $\Gamma=\alpha\gamma_\parallel$ from the experimental data, and fitting them to the numerically extracted values, we get a value for $\gamma_\parallel/2\pi \approx \SI{6.25e-4}{\hertz}$ ($T_1\approx\SI{30}{min}$).
\\\\\textbf{Theoretical modelling}\\
To calculate the quench dynamics displayed in Fig.~\ref{fig:quench}, we numerically solve the Maxwell-Bloch equations (Eqs.~(\ref{eq:maxwellbloch})) for the driving signals chosen nearby the first-order transition (see the main text for details) using the standard Runge-Kutta method. As an initial condition we take the steady state given by Eqs.~(\ref{eq:a2}), which lies on the upper branch depicted in Fig.~\ref{fig:bistab}c and corresponds to the limit of strong driving with $|\sigma_j^z|,\, |\sigma^-_j|\ll 1$. To accurately describe the dynamics and to achieve a good correspondence with experimental data, we take into account the effect of an inhomogeneous broadening by modelling the spin density with a $q$-Gaussian shape for the spin density, $\rho(\omega)=C\cdot\left[1-(1-q)(\omega-\omega_s)^2/\Delta^2\right]^{1/(1-q)}$, distributed around the mean frequency $\omega_s/2\pi=2.6915$\,GHz with the parameter $q=1.39$ and the width $\Delta/2\pi=5.3$\,MHz. Such a shape for $\rho(\omega)$ was previously established by obtaining an excellent agreement between our theoretical model and the experiment, when treating the problem in the framework of the Volterra equation valid in the limit of weak driving signals \cite{krimer_non-markovian_2014}. We then straightforwardly discretize our problem by performing the following transformation, $g_j=\Omega \cdot \left[\rho(\omega_j)/\sum_l\rho(\omega_l)\right]^{1/2}$, where $\Omega^2=\sum_j g_j^2$ stands for the collective coupling strength ($\Omega/2\pi=9.6$\,MHz and $12.6$\,MHz  for Figs.~\ref{fig:bistab}ab and Fig.~\ref{fig:bistab}c, respectively). Since in total we deal with a sizable number of spins ($N\approx 10^{12}$), we make our problem numerically tractable by dividing spins into many subgroups with approximately the same coupling strengths, so that the numerical values for $g_j$ in Eq.~(\ref{eq:maxwellbloch}) represent a coupling strength within each subgroup rather than an individual coupling strength.\\
\\\textbf{Adiabatic Elimination}\\
Using the fact that $\gamma_{\parallel} \ll \kappa,\gamma_{\perp}, \Omega$, the dynamics at large times when $t \gg 1/\kappa,1/\gamma_{\perp}, 2\pi/\Omega$ can be considerably simplified as the cavity amplitude $a$ and the spin lowering expectation values $\sigma^-_{j}$ follow adiabatically the evolution of the $z$-component of the spin operator expectation value, $ \sigma_z^j$. By introducing the small parameter $\epsilon=\gamma_{\parallel}/\gamma_{\perp}$ and the slow dimensionless time $\tau=\gamma_{\parallel} t$, we finally derive a first order differential Eq.~(\ref{eq:adiabatic}) for the intra-cavity intensity $|a|^2$.\\
}

\FloatBarrier

\bibliography{main.bib}

\end{document}